# Electronic structure of oxygen functionalized armchair graphene nanoribbons

Adam J. Simbeck[1‡], Deyang Gu[1,2], Neerav Kharche[1,3,4], Parlapalli Venkata Satyam[5], Phaedon Avouris[6], and Saroj K. Nayak[1,7]

[1] *Department of Physics, Applied Physics, and Astronomy, Rensselaer Polytechnic Institute, Troy, NY, 12180, USA*
[2] *Department of Computer Science, Rensselaer Polytechnic Institute, Troy, NY 12180, USA*
[3] *Computational Center for Nanotechnology Innovations, Rensselaer Polytechnic Institute, Troy, NY 12180, USA*
[4] *Chemistry Department, Brookhaven National Laboratory, Upton, NY 11973, USA*
[5] *Institute of Physics, Sachivalaya Marg, Bhubaneswar-751005, India*
[6] *IBM Research Division T. J. Watson Research Center, Yorktown Heights, NY 10598, USA*
[7] *School of Basic Sciences, Indian Institute of Technology Bhubaneswar, Bhubaneswar-751013, India*

The electronic and magnetic properties of varying width, oxygen functionalized armchair graphene nanoribbons (AGNRs) are investigated using first-principles density functional theory (DFT). Our study shows that O-passivation results in a rich geometrical environment which in turn determines the electronic and magnetic properties of the AGNR. For planar systems a degenerate magnetic ground state, arising from emptying of O lone-pair electrons, is reported. DFT predicts ribbons with ferromagnetic coupling to be metallic whereas antiferromagnetically coupled ribbons present three band gap families: one metallic and two semiconducting. Unlike hydrogen functionalized AGNRs, the oxygen functionalized ribbons can attain a lower energy configuration by adopting a non-planar geometry. The non-planar structures are non-magnetic and show three semiconducting families of band gap behavior. Quasiparticle corrections to the DFT results predict a widening of the band gaps for all planar and non-planar, semiconducting systems. This suggests that oxygen functionalization could be used to manipulate the electronic structures of AGNRs.

## I. INTRODUCTION

Since the synthesis and characterization of graphene[1] the search has been ongoing to uncover a viable method to take advantage of its massless, relativistic charge carriers (Fermi velocity up to $10^6$ m/s) and high mobility [up to 15,000 cm$^2$/(Vs) when supported on SiO$_2$, and 200,000 cm$^2$/(Vs) or greater when supported on BN or free-standing, with a current record of ~1,000,000 cm$^2$/(Vs)].[1-7] An attractive option to modulate the electronic properties of graphene is to cut the two-dimensional (2D) sp$^2$-hybridized sheet of C atoms to form a quasi-one-dimensional (1D) ribbon, a graphene nanoribbon (GNR). The electronic properties of these carbon allotropes depend strongly on their edge morphology, which can be armchair, zigzag, or intermediate.[8,9] Currently, a number of experimental techniques are in development to grow GNRs with atomistically precise edge structure.[10] In addition the dangling σ bonds of the unpassivated edge C atoms present another unique opportunity to tune the electronic properties of the GNRs. Both armchair graphene nanoribbons (AGNRs) and zigzag graphene nanoribbons (ZGNRs) have undergone intense theoretical and experimental investigation over the past two decades and show promise for use as both active and passive components in electronic circuits.[11-26]

It has been theoretically shown that H-passivated AGNRs (H-AGNRs) are all semiconductors which belong to one of three families: *3p, 3p+1,* or *3p+2* (where *p* is an integer).[11,17] Similarly, H-passivated ZGNRs (H-ZGNRs) also behave as semiconductors, but whereas the emergence of a band gap in H-AGNRs is attributed to quantum confinement and edge effects, the band gaps in H-ZGNRs result from a staggered sub-lattice potential arising from magnetic ordering near the edges.[11] Both H-AGNRs and H-ZGNRs show band gaps that decrease as a function of increasing width, and these trends have been confirmed by more accurate quasiparticle (*GW*) calculations.[27] Ref. 27 also highlights that for GNRs, *GW* calculations generally increase the band gaps predicted by density functional theory (DFT) without altering the ordering of the families. Besides H-passivation, the effects of a number of other edge passivations have been considered.[15,19,21,25,28] In general, the resultant electronic structure is highly dependent upon the type, concentration, and edge geometry of the passivation species.[14,19,21] One of the more interesting choices to saturate the dangling σ bonds is oxygenation[12-15,18-22,26] since O is readily present given the experimental methods to fabricate GNRs, such as O-plasma etching.[29-33] Furthermore, it has been theoretically shown that oxygenated GNRs are more stable than hydrogenated ones.[14]

O-passivated ZGNRs have been the focus of a variety of studies because of the inherent magnetic behavior of the bare ZGNR.[14,15,19,22,26,34] Although bare AGNRs lack such spin-polarization effects, they too have been widely studied since it has been theoretically shown that AGNRs are more stable than ZGNRs.[16,35] Two levels of concentration for AGNRs with O-passivation (O-AGNRs) are usually considered: one in which each O atom bonds to two edge C atoms (2 O atoms per unit cell), sitting within the so-called *armchair* of the edge, and one in which each O atom is bonded to only a single edge C atom [4 O atoms per unit cell (Fig. 1)]. The first type of edge geometry has been shown to drastically modify the electronic properties of AGNRs but does not show spin-polarization effects.[12,13,15,21]

The second type of concentration, i.e. 4 O atoms per unit cell such that each O atom bonds to a single edge C atom, has been studied considerably less. Note that from here on out, unless otherwise explicitly stated, the shorthand O-AGNR shall refer to only this second concentration. Seitsonen *et al.*, theoretically showed that such bonding results in the most stable O-passivated AGNR, not including edge substitution.[15] Here a non-planar, nonmagnetic (NM), and non-metallic ground state is found. Hu *et al.*, also studied such oxygenation and the behavior of O-AGNRs under an applied electric field using DFT.[18] Here a planar, metallic ground state is reported, in contrast to Ref 15. Spin-polarization effects were not reported in Ref. 15 or 18. The metallic behavior in the planar configuration is attributed to impurity states at the Fermi energy from O lone-pair electrons and π electrons from the GNR interior C atoms. Furthermore, it is shown that the electronic structure of the O-AGNR is sensitive to the strength of the applied electric field and can result in semiconducting-to-metallic transitions. Finally, this type of edge oxygenation has also been studied by Pan and Yang.[20] Here it is reported that passivation of each edge C atom with an O atom results in a planar, NM semiconductor with a band gap of 0.4 eV, in contrast to Refs. 15 and 18. The NM ground state, a result reported to be consistent with Ref. 15, is attributed to the lack of localized states formed at the Fermi energy. Since the fully oxygenated case is uninteresting from a magnetic viewpoint, the concentration of O atoms is reduced. This in turn induces a ferromagnetic (FM) behavior in which the spin-polarizations of the unpassivated edge C and O atoms are aligned. The structure is then found to be a semimetal and the magnetism is ascribed to an electron transfer between edge C and O atoms.

[‡] Corresponding author

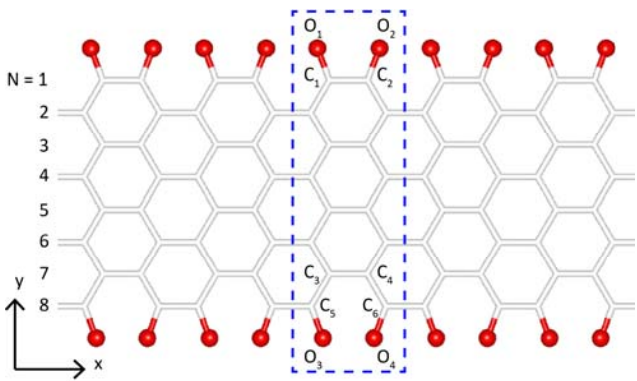

FIG. 1. (Color online) Geometry of planar O-AGNR with a width of $N = 8$. The width of the unit cell (blue, dashed rectangle) is taken as 4.26 Å. The labeling convention for the C (white) and O (red) atoms is for bonding and partial density of states (PDOS) analysis.

The works outlined above expose the fact that the understanding of the electronic and magnetic character of O-AGNRs is both incomplete and in some instances inconsistent. For example, what is the reason behind the disagreement in the previous references on the ground state configuration? Such a question must be answered using the same methodology (i.e. same scheme, code, pseudopotential, etc.) in order to provide confidence in any energetic predictions. The literature also lacks more accurate band gap predictions using the *GW* method, which in the case of H-AGNRs has been shown to widen gaps by as much as 2 eV compared to DFT results.[27] Lastly, a study of the width dependence of the electronic and magnetic properties of O-AGNRs is also absent, even though the H-AGNR case again highlights its importance.[11] These are the type of issues to be addressed in this paper.

In this study, a first-principles DFT method is employed to analyze the electronic/magnetic properties of O-AGNRs of increasing width. It is found that the environment of the edge geometry is rich with metastable configurations and the most stable planar and non-planar systems are analyzed. The geometry of the O-AGNRs in turn determines their electronic and magnetic behavior. The planar systems are found to have spin-polarized ground states which display FM coupling between atoms at each edge and either FM or antiferromagnetic (AFM) coupling between the two edges. The systems with FM coupling between the edges are designated as FM O-AGNRs and those with AFM coupling as AFM O-AGNRs. The DFT calculations predict FM O-AGNRs to always be metallic, independent of width, whereas the AFM O-AGNRs fall into three families, similar to H-AGNRs except that the *3p* family is predicted to be metallic. Computationally more expensive many-body perturbation theory (MBPT) calculations within the *GW* approximation are also carried out to verify such trends. Here, the band gaps of the semiconductor-like systems are widened. Finally, non-planar O-AGNRs are also investigated where a ~2 eV lowering of energy, as compared to the planar geometry, is reported. The non-planar systems exhibit two stable geometries: one in which O atoms on the same side of the unit cell but opposite edges are both displaced along the same direction perpendicular to the ribbon plane, and another where such atoms are opposed from one another (Fig. 7). Both non-planar O-AGNR geometries yield a NM, semiconducting ground state with band gaps that decrease as a function of increasing ribbon width. These results can be used to interpret those previously reported in Refs. 15, 18 and 20.

## II. METHOD

For the calculations of the electronic and magnetic properties of the planar (Fig. 1) and non-planar (Fig. 7) O-AGNRs the DFT based Vienna *ab initio* simulation package (VASP)[36-39] was used. Projector-augmented wave (PAW) pseudopotentials[40, 41] were employed to describe the electron-ion interaction. The exchange-correlation energy is described within the local-spin density approximation (LSDA)[42] and similar results are found using the generalized gradient approximation (GGA).[43, 44] Furthermore, the use of ultrasoft pseudopotentials[45, 46] is also found to not alter any of the predictions. The energy cutoff is set at 500 eV and at least 15 Å of vacuum is included along directions perpendicular to the periodic direction of the O-AGNR to prevent any interaction between periodic images. The width of the unit cell (Fig. 1) is assumed to be that of a pristine ribbon, i.e. 4.26 Å. Such a choice results in a pressure on the unit cell of only a few kB. For the geometry optimization, a Monkhorst-Pack *k* point grid of $32 \times 1 \times 1$ is employed to sample the 1D Brillouin zone. The positions of the ions are relaxed until the force on each atom is less than 0.01 eV/Å and the difference in total energy is less than $10^{-4}$ eV. Gaussian smearing with a width of 0.05 eV is used to describe the partial occupancies of orbitals.

For the quasiparticle corrections to the band gaps MBPT is employed within the *GW* approximation in the ABINIT code.[47] Norm-conserving pseudopotentials are generated using the Trouiller-Martins scheme[48] and the Teter-Pade[49] approximation is employed to describe the exchange-correlation energy. The quasiparticle corrections are calculated within the $G_0W_0$ approximation, and screening is calculated using the plasmon-pole model.[50] To ensure negligible interaction between periodic images, 10 Å of vacuum is included in directions perpendicular to the periodic direction of the nanoribbon. The sampling of the Brillouin zone is identical to that used for the calculations in VASP. For all *GW* calculations the Coulomb cutoff [51] is taken into consideration.

## III. RESULTS

The results are presented as follows. First the planar O-AGNR structures in the NM state are discussed in order to motivate trends in the geometry, electronic band structure, and partial density of states (PDOS). Next, findings on the magnetic systems are presented, where again band structure and PDOS figures, as well as plots for the spin density, are shown. Then the electronic structure of the most energetically favorable systems, the non-planar O-AGNRs, is discussed. Finally the results of this work are compared to those of Refs. 15, 18 and 20. Essentially the O-AGNRs have a number of metastable geometries that only significantly differ at the edges, which in turn leads to the various contradicting results presented in the aforementioned references.

### A. Nonmagnetic O-AGNRs

The geometry for all planar structures is featured in Fig. 1. Although for all magnetic initializations the geometry is allowed to fully optimize, the NM, FM, and AFM systems all converge to identical geometries. The C-C bonding in the middle of the ribbon agrees with that of pristine graphene, i.e. ~1.42 Å. Near the edges of the nanoribbon though, the ideal C-C bonding is disrupted by the O-passivation. For example, for the O-8-AGNR the $C_3$-$C_4$ bond becomes 1.40 Å and the $C_1$-$C_2$ bond is 1.44 Å. Note that the GNR of width N has been labeled as O-N-AGNR. This C-C bond disruption is not as severe as in the case of H-passivation.[11] The C-O separation is found to have a strong double bond character at 1.26 Å. The separation between O atoms in the unit cell is $O_1$-$O_2$ = 2.34 Å and $O_3$-$O_4$ = 1.92 Å. These distances are large enough to conclude that the O atoms do not bond to one another. Finally the $O_1$-$C_1$-$C_2$ angle is 111° and the $O_3$-$C_5$-$C_3$ angle is 129°.

Here it is noted that another planar geometry, in which the O atoms pair up, is also found but is ~0.7 eV per unit cell higher in energy than the structure previously described. The bonding and angles of this less energetically favorable geometry are as follows: $C_3$-$C_4$ = 1.43 Å, $C_1$-$C_2$ = 1.41 Å, C-O = 1.37 Å, $O_1$-$O_2$ = 2.76 Å, $O_3$-$O_4$ = 1.50 Å, $O_1$-$C_1$-$C_2$ = 120°, and $O_3$-$C_5$-$C_3$ = 120°. Therefore the planar system is rather sensitive to the edge geometry, specifically the C-O bond length, the separation between O atoms in the unit cell, and obviously then the O-C-C bond angle.

The band structure and PDOS of the NM O-AGNR system for a width of $N$ = 8 is featured in Fig. 2. The middle panel shows the PDOS due to each of the O atoms and the right panel is the PDOS of the edge C atoms ($C_1$, $C_2$, $C_5$, and $C_6$ in Fig. 1). The C-O bonding in the AGNR case should be similar to that seen in O-ZNGRs.[14, 26] More specifically, it is expected that the dangling $sp^2$-σ bond of the edge C atoms will be passivated by O. Furthermore, the single electron in the $p_z$ orbital of the O atom is expected to form a π bond with the edge C atom. The prediction of a double bond is affirmed by the small C-O bond length of 1.26 Å. Lastly the bands nearest the Fermi energy $E_F$ should arise from the non-bonding lone-pair electrons of the O atoms.

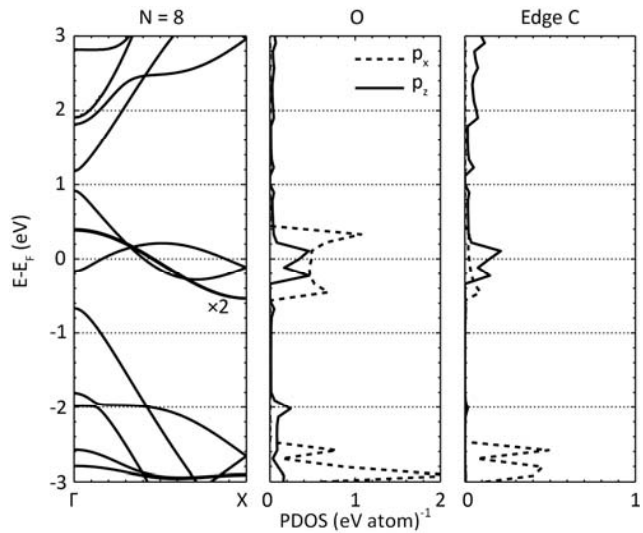

FIG. 2. Band structure and PDOS of NM O-8-AGNR. The spin-degenerate band structure (left) shows four bands near $E_F$ which can be attributed to the O (middle) lone-pair and $p_z$ electrons, and to a lesser extent edge C (right) atoms. Edge C here refers to $C_1$, $C_2$, $C_5$, and $C_6$ in Fig. 1.

From Fig. 2 it is seen that the characteristic π and π* bands from the C atoms of the AGNR are preserved and appear just above ±1 eV. Unlike the H-passivated case though, O introduces four bands, two of which are degenerate, that each cross the Fermi level, resulting in metallic behavior, and widens the π-π* gap from 0.26 eV to 1.85 eV. Of the four bands, the two degenerate ones are ascribed to O-$p_x$ lone-pair electrons, whereas the other two arise from O-$p_z$ states. This is confirmed by direct inspection of the charge density of these four bands, as well as the PDOS analysis (Fig. 2, middle). Comparing the PDOS over the full energy range also reveals a strong hybridization between the O-$p_y$ and edge C-$p_y$ states, indicating the formation of a σ bond, as well as the mostly passivated edge C-$p_z$ state to form a π bond with the O-$p_z$ state. The bonding picture for O-AGNRs is markedly different from the simple analysis previously described. A double bond between C and O is still formed, but the lone-pair bands are partially emptied to fill O-$p_z$ states near the Fermi level. A similar partial emptying (filling) of lone-pair ($p_z$) bands is seen in O-ZGNRs and sulfur passivated ZGNRs (S-ZGNRs).[14, 26] For O-AGNRs though, the lone-pair band emptying is more severe since the O atoms are closer to one another, and in general appears to be rather sensitive to the O atom spacing. Furthermore, the counterintuitive bonding picture for O-AGNRs is an indication of an instability with the planar ribbon geometry as a result of the concentration of O in the unit cell, which is not the case for O-ZGNRs. This point is discussed more in the non-planar geometry section (III.D). Since the bands crossing $E_F$ arise from O lone-pair electrons and $p_z$ states, increasing the width of the ribbon will not affect the metallic behavior. Finally, it is pointed out that for an 11-O-AGNR, the planar, nonmagnetic results agree with Ref. 18.

## B. Ferromagnetic O-AGNRs

The inclusion of spin-polarization leads to an O-AGNR in which there is FM ordering along each edge and FM coupling between the two edges (Fig. 3). The FM O-8-AGNR is found to be 0.23 eV per unit cell lower in energy than the NM O-8-AGNR. For longer ribbons the energy difference between the NM and FM states converges to 0.22 eV per unit cell. The magnetization density, i.e. the difference between the spin up and spin down densities $\rho_\alpha$-$\rho_\beta$, is plotted in Fig. 3. The top plot [Fig. 3(a)] is the magnetization density integrated along the z-direction, hence the units of e/Å$^2$, and a contour plot of the magnetization density is featured below it [Fig. 3(b)]. Both figures highlight that the magnetization is mainly confined to O and the first two rows of edge C atoms. Similar to the case of graphone,[52] the FM ordering along each edge is attributed to the larger spatial extent of p$_z$ electrons which promotes long-range exchange-coupling interactions. For FM O-8-AGNR the magnetic moment per unit cell is 3.4 $\mu_B$ and per O atom is 0.54 $\mu_B$. Both of these values converge as the ribbon width is increased: total magnetization to 3.7 $\mu_B$ per unit cell, and magnetization per O atom to 0.55 $\mu_B$. A similar magnetic moment convergence trend is reported in Ref. 20 for AGNRs passivated with a lesser concentration of O.

The band structure and PDOS for the FM O-AGNR of width $N = 8$ is given in Fig. 4. Although the overall general shapes of the bands are consistent with the NM system, the bands are no longer spin degenerate and have shifted position. The spin splitting is most noticeable for the bands near $E_F$, especially for the nearly flat bands which envelope those that cross the Fermi level. A similar strong spin-polarization for lone-pair bands of O-AGNRs is reported in Ref. 20. As in the NM case, the eight bands near $E_F$ are considered to be primarily from the lone-pair and p$_z$ electrons of O. Here the PDOS shows that the exchange splitting is more severe for the p$_x$ lone-pair states. Also appearing is a segregation of p states in which the p$_z$ states lie closer to $E_F$ and are essentially bounded by the p$_x$ states. This trend will be seen again in the AFM systems. It is also interesting to note that out of the eight bands nearest $E_F$, the four spin-down states are almost completely empty, whereas the spin-up states are nearly full, giving rise to the FM behavior. The O-p$_x$ states are now either completely full (spin-up) or completely empty (spin-down), whereas the p$_z$ states are either mostly full (spin-up) or mostly empty (spin-down), in contrast to the NM state where both states were partially filled. Essentially the emptying (filling) of the p$_x$ lone-pair electrons (p$_z$ electrons) has become more complete for a given spin state, hence the lowering of energy. Similar to the NM case the π and π* bands of the nanoribbon are preserved and the π-π* gap is again increased: 1.76 eV for spin-up electrons and 1.84 eV for spin-down electrons. Finally, since the metallic and magnetic nature is still confined to the edges, increasing the width of the ribbon will not affect such trends.

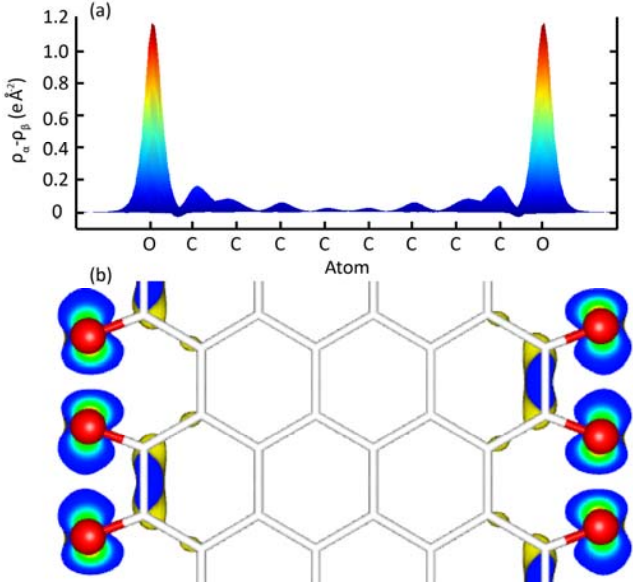

FIG. 3. (Color online) (a) Integrated and (b) contour magnetization density plot for FM O-8-AGNR. Both show that the spin is concentrated mainly on the O atoms, but also slightly extends into the edge C atoms.

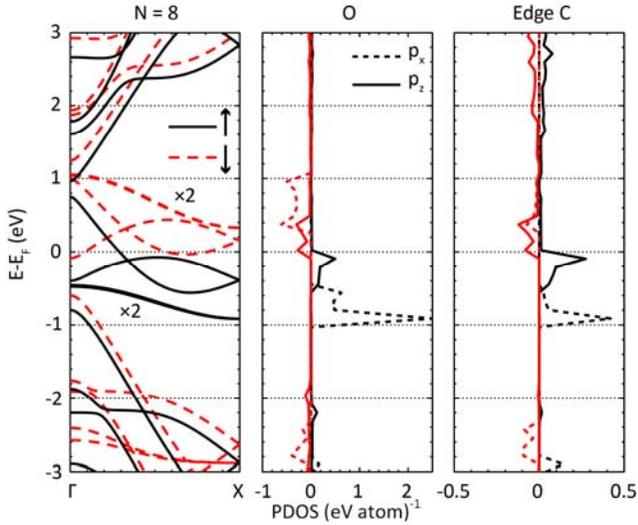

FIG. 4. (Color online) Band structure and PDOS of FM O-8-AGNR. The band structure (left) shows that the spin up and spin down bands are non-degenerate. The PDOS shows that the bands near $E_F$ are mainly due the p$_x$ and p$_z$ states of the O atoms (middle). Note that the spin up states (black/dark) are plotted along the positive axis whereas the spin down states (red/light) are plotted along the negative axis.

## C. Antiferromagnetic O-AGNRs

A second magnetic state, which is as stable as the FM one, is found where again there is FM ordering along each edge but now the coupling between edges is AFM. The AFM state for O-8-AGNR is 30 meV per unit cell lower in energy than the FM state. The energy difference between the AFM and FM states converges to zero as the ribbon width is increased. For example, the difference in energy between the two magnetic states for an O-AGNR of width $N = 29$ is 0.8 meV per unit cell. The magnetic density appears identical to that featured in Fig. 3, except that now one edge is polarized along the opposite direction. The magnetic moment per unit cell is obviously zero and the magnetic moment per O atom is consistent with that found in the FM case. For example, the magnitude of the magnetic moment per O atom for the AFM O-8-AGNR is 0.54 $\mu_B$. Furthermore, the magnitude of the magnetic moment per O atom again converges to 0.55 $\mu_B$ as the width of the nanoribbon is increased.

The band structures and PDOS of the planar, AFM O-AGNR systems are shown in Fig. 5. Similar to the case of H-passivation,[11] three families for the energy gap trends are found: $3p$, $3p+1$, and $3p+2$, where $p$ is an integer. In Fig. 5 the electronic structure of a representative member of each family, namely $N = 6$, 7, and 8, is shown. DFT calculations predict that the $3p+1$ and $3p+2$ families are both semiconductors with band gaps that decrease as the width of the ribbon is increased, whereas the $3p$ family is predicted to be metallic (Fig. 6). Note that the definition of the nanoribbon width here is identical to that of Ref. 11. Again for all families four (eight, including spin) bands are found near the Fermi level. The PDOS for each of the ribbons shows that the major contributions are still attributed to the O lone-pair electrons ($p_x$ states), but now a lone-pair band has been completely emptied to fill an O-$p_z$ state. Unlike the NM and FM states though, only O atoms on the same edge give identical DOS contributions. Rather for AFM O-AGNRs, similar to the case of bare ZGNRs,[14] the edge localization of the states associated with the O-$p_x$ lone-pair and O-$p_z$ bands is dependent upon the spin state. For instance, the filled, spin up (down) O-$p_x$ states are localized at the $O_{3,4}$ ($O_{1,2}$) atom edge, and vice versa for the empty bands. This gives rise to the net spin magnetic moment of zero. It should also be pointed out that the $p_z$ states of the O atoms are always found closer to $E_F$ and it is the shifting of such states towards $E_F$ that eventually causes the semiconducting families to become metallic at large enough widths (Fig. 6). In other words, the positioning of the flatter bands, i.e. the second closest bands above and below $E_F$ which are from the O-$p_x$ lone-pair electrons, is relatively stable as the width of the nanoribbon is increased. The positions of the bands closest to the Fermi level on the other hand are not. The enveloping nature of the $p_x$ states is consistent with the FM O-AGNR results. Lastly, the preservation of the characteristic $\pi$ and $\pi^*$ bands from C is still seen, as well as an increased $\pi$-$\pi^*$ gap as compared to H-AGNRs.

From Fig. 6 it is seen that the energy gaps of all semiconducting nanoribbons follow a 1/width dependence, except for $N = 4$. Rather from $N = 4$ to $N = 7$, which are both members of the $3p+1$ family, there is an increase in the band gap from 0.35 eV to 0.42 eV. This trend was confirmed using the GGA to describe the exchange-correlation energy. It is suspected that such a trend is a result of the small width (3.64 Å) of the $N = 4$ AFM O-AGNR, which allows the two edges of the ribbon to interact and reduce the energy gap. For example, Fig. 3 shows that the magnetization is confined to the O atoms and the first two rows of C atoms, which for an $N = 4$ ribbon is the entire width. Similar deviations for very small width O-AGNRs are also seen in the non-planar results (Fig. 9). Fig. 6 also features the $GW$ corrections to the energy gaps for $N = 5$ and 7. As expected the energy gaps in the semiconducting systems are increased and the $GW$ corrections do not change the ordering of the families in terms of the band gap trends.

As compared to H-AGNRs the band gap for each family in Fig. 6 is generally reduced,[11] even with the $GW$ corrections.[27] Furthermore, the ordering of the families in terms of the band gap trends is different than the case of H-passivation. For

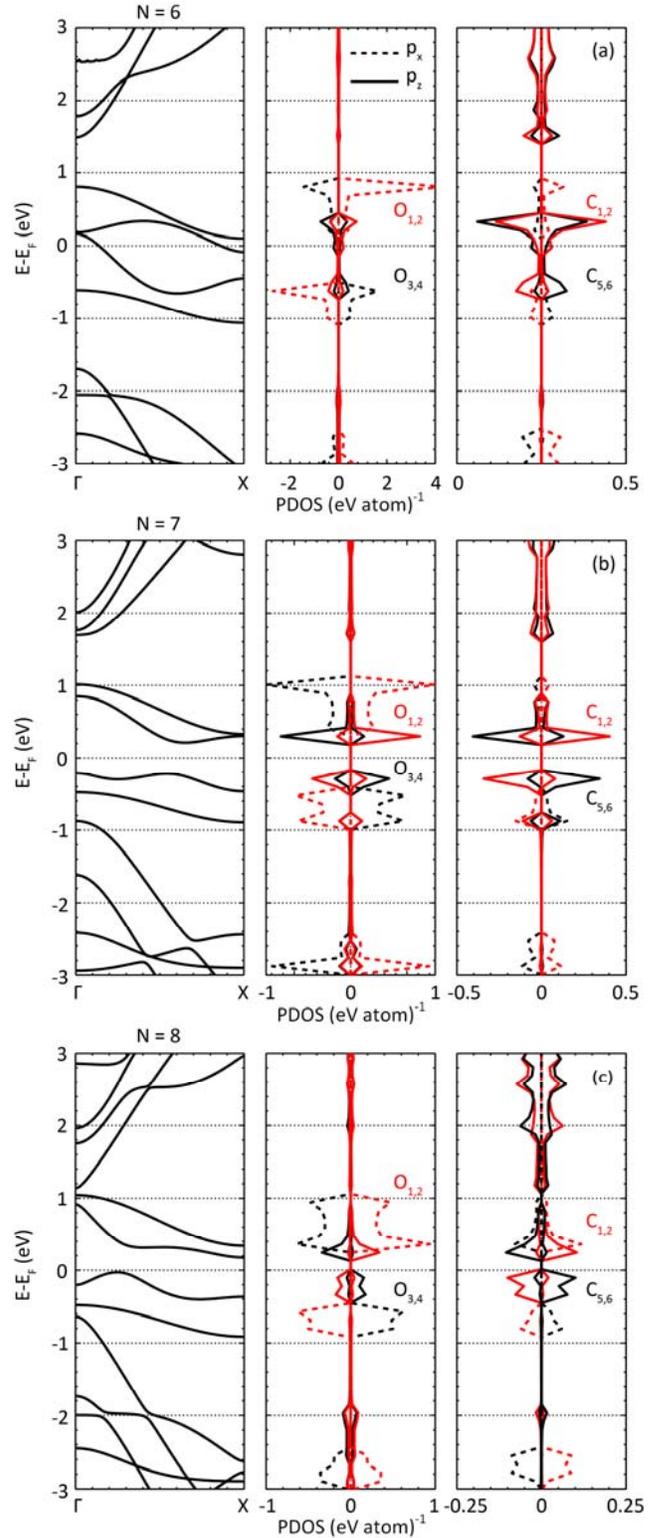

FIG. 5. (Color online) Band structure and PDOS of the three AFM O-AGNR families. The band structure (left) for each family is spin degenerate. (a) $N = 6$ ($3p$) is a metal, whereas (b) $N = 7$ ($3p+1$) and (c) $N = 8$ ($3p+2$) are semiconductors with indirect band gaps of 0.42 eV and 0.20 eV respectively. The PDOS is again dominated by the O (middle) lone-pair and $p_z$ electrons but the symmetry of Fig. 2 is lost. Note that spin down contributions are plotted along the negative axis.

AFM O-AGNRs the ordering is

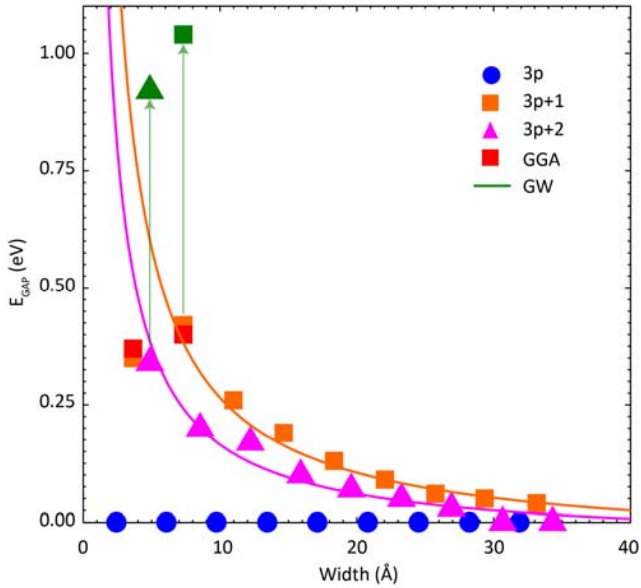

FIG. 6. (Color online) Energy gap trends for the three AFM O-AGNR families. DFT predicts that the *3p* family is always metallic. DFT also predicts a gap that decreases with width for the *3p+1* and *3p+2* families. *GW* calculations (green) widen the band gaps in the semiconducting systems (*3p+1*, *3p+2*). The lines have been added to emphasize the ~1/width dependence. The arrows show the GW corrections to the band gap for *N* = 5 and 7.

whereas for H-passivation it is $E_{Gap}^{3p+2} < E_{Gap}^{3p} < E_{Gap}^{3p+1}$.[11] The family ordering is also different than that of ribbons passivated with a lower concentration of O atoms (2 O atoms per unit cell) such that each O atom is bonded to two C atoms and sits within the *armchair* of the edge. The ordering for these systems is $E_{Gap}^{3p+1} < E_{Gap}^{3p+2} < E_{Gap}^{3p}$.[13] It is interesting to note that with this reduced concentration of O, as compared to H-passivation, the gap is generally increased, except in the *3p+1* family, whereas in the case of AFM O-AGNRs the gap is generally decreased.

### D. Non-planar O-AGNRs

Finally the findings for the non-planar O-AGNRs are presented. Figure 7 shows a side view of the two stable non-planar geometries. The first structure is one in which the O atoms on opposite edges but same sides of the unit cell are displaced from the plane of the ribbon in the same direction [Fig. 7(a)], and hence in the side view only half the O atoms are fully visible. This structure is referred to as $O_{SAME}$-AGNR. In the other stable structure these O atoms are displaced in opposite directions from the nanoribbon plane [Fig. 7(b)]. This system is referred to as $O_{OPP}$-AGNR. Both geometries are more energetically favorable than the planar one, but energetically comparable to one another. For example, the energy difference between the planar AFM O-8-AGNR and non-planar, NM O-8-AGNR is 1.94 eV and converges to ~2 eV per unit cell as the ribbon width is increased. The energy difference between $O_{SAME}$-8-AGNR and $O_{OPP}$-8-AGNR on the other hand is merely 2 meV per unit cell and converges to zero as the nanoribbon width increases. There exists no appreciable reaction barrier for the system to transition from the planar to the non-planar structures,[54] and therefore the non-planar geometries should indeed be the ground state whereas the planar systems appear to be merely hypothetical. The existence of a more energetically favorable non-planar structure is consistent with Ref. 15.

Besides the displacement of the O atoms, the geometries of the two non-planar structures are identical to one another. For instance, the C-O bond for both non-planar systems is 1.21 Å. As compared to the planar O-AGNRs, the interior geometry of the ribbons remains intact but the edges are obviously distorted. For example the $C_1$-$C_2$ and $C_3$-$C_4$ bonds are 1.51 Å and 1.38 Å respectively. In addition, the $C_3$-$C_5$ bond is increased from 1.42 Å to 1.48 Å. As the O atoms move away from the plane of the nanoribbon they drag along with them the edge C atoms which gives rise to such changes in the edge C-C bonding (Fig. 7). It is suspected that the transition to a non-planar geometry in the presence of oxidation is to accommodate the steric interaction between O atoms. Essentially, the non-planar geometry allows the O atoms to move further apart from one another (2.75 and 2.96 Å) than what can be achieved with a planar system (1.92 and 2.34 Å). The non-planar geometry also allows for a more efficient hybridization of atomic orbitals (Fig. 8). In O-ZGNRs, the separation between O atoms is already large (2.46 Å) so that there is no need for neighboring passivations to assume opposite tilts with respect to the nanoribbon plane.[26]

In addition to the geometries of the non-planar systems being identical to one another, the electronic structure of $O_{SAME}$-AGNR and $O_{OPP}$-AGNR is also similar for ribbons with N ≠ 4 (Fig. 8 and 9). Again three families of behavior for the band gap are found, which are all semiconductors, except for N = 3 and $O_{OPP}$-4-AGNR (Fig. 9). The PDOS again shows that the bands near $E_F$ are primarily from O, with minor contributions from edge C atoms. Note that unlike the previous PDOS though, the $p_y$ states are included in Fig. 8 since they now show contributions comparable to the $p_x$ and $p_z$ states. Note that the O-$p_x$ states have been scaled by 0.2 in Fig. 8. The usual four bands near the Fermi level are now split into two groups: one set above $E_F$ and attributed to O-$p_z$ states, and the other set below $E_F$ and attributed to O lone-pair electrons. The non-planar geometry allows the lone-pair states to be completely filled and O-$p_z$ states to be completely empty, as expected from the previous naïve bonding analysis. Also note that these four bands are still segregated from the characteristic π and π* states of the interior C atoms, which begin to appear below -0.5 eV and above 1.5 eV, and the π-π* gap is again increased to 2.29 eV. Finally, the PDOS shows that the correlation between all p-like states is strongest in the non-planar geometry. This suggests that a more efficient hybridization, and hence a lowering of energy, can be achieved when the O and edge C atoms are allowed to move out of the nanoribbon plane.

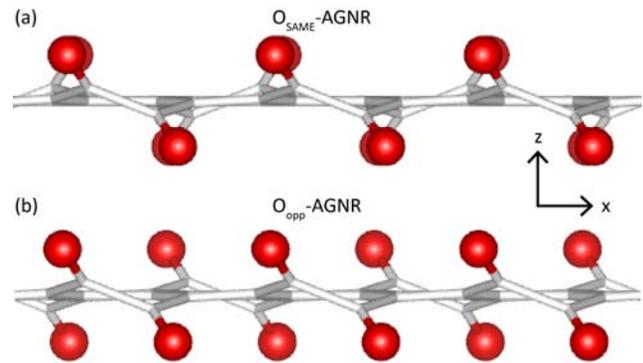

FIG. 7. (Color online) Geometry of (a) $O_{SAME}$-AGNR and (b) $O_{OPP}$-AGNR. The steric interaction between edge O atoms disrupts the planar geometry of the nanoribbon leading to a more stable, non-planar structure.

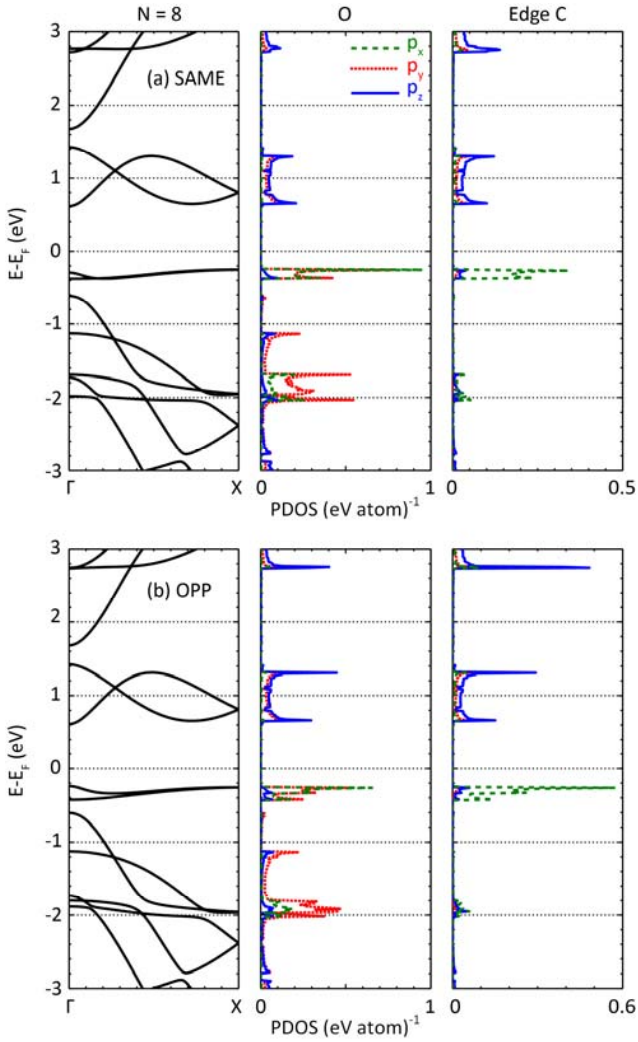

FIG. 8. (Color online) Band structure and PDOS of non-planar O-8-AGNR. The two stable geometries, i.e. (a) $O_{SAME}$-8-AGNR and (b) $O_{OPP}$-8-AGNR, show similar band structures (left) and PDOS. Furthermore, the PDOS near $E_F$ is again dominated by the p states of O (middle). Note that the O atoms $p_x$ states have been scaled by 0.2.

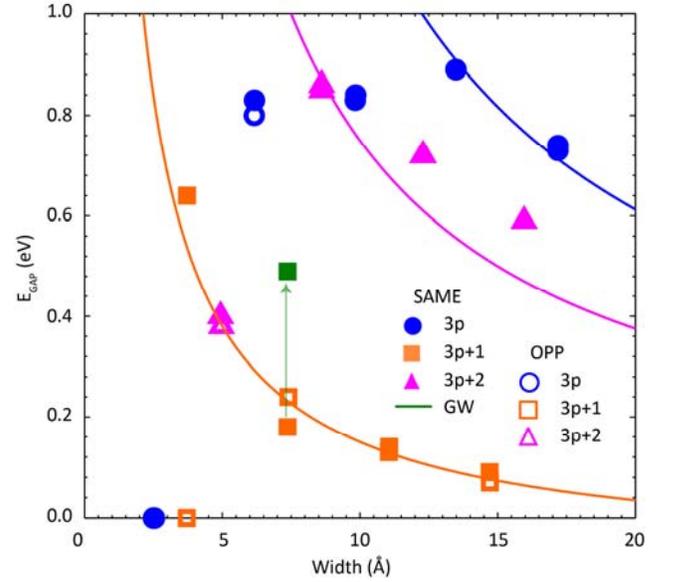

FIG. 9. (Color online) Energy gap trends for the three non-planar O-AGNR families. DFT again predicts a gap that decreases with width for all three families. The *GW* calculation (green) again widens the gap. The lines have been added to emphasize the ~1/width dependence. The arrow shows the GW correction for the $O_{SAME}$-7-AGNR.

From Fig. 9 the band gaps for each family generally follow the 1/width dependence and any deviations from such can be traced back to the small width of the non-planar O-AGNR. For instance, DFT predicts the non-planar $N = 3$ ribbon to be metallic. Such a ribbon though is completely non-planar since the width is spanned by merely three dimer lines of C atoms, all of which move away from the nanoribbon plane with the O atoms. This drastic contortion of the non-planar O-3-AGNR gives rise to two bands crossing $E_F$ and hence metallic behavior. By increasing the width of the non-planar O-AGNR for each family though, good agreement is achieved with the 1/width dependence. Here it is noted that since DFT generally predicts the non-planar O-AGNRs to be semiconducting, it is anticipated that *GW* calculations will only increase the band gap for such cases and is also not expected to alter any band gap family trends or ordering. In Fig. 9 a single *GW* band gap correction is shown for $O_{SAME}$-7-AGNR to confirm this hypothesis.

As compared to the planar geometry band gap trends, not only is a reordering of the families ( ) found but also a general increase in the band gaps, except for the *3p+1* family. Comparing to H-AGNRs, a reordering of the families is again found, but only the 3p (beyond N = 9) and *3p+2* families show an increased band gap, whereas the *3p+1* family shows a decrease. This is identical to the trend seen with a reduced concentration of oxgyen atoms.[13] This range of results highlights the fact that even when considering only a single type of passivation species, the concentration and edge geometry can have a drastic effct on the magnetic and electronic stucture.

## IV. DISCUSSION

Here the above results are discussed in light of those previously reported in Refs. 15, 18, 20. Ref. 20 reported that O-AGNRs are planar, NM semiconductors. More specifically it was predicted that an O-11-AGNR has a band gap of 0.4 eV and that the NM result is consistent with Ref. 15. For a planar O-11-AGNR a stable FM and AFM ground state, which are 0.24 eV per unit cell lower in energy than the NM state and have only an 8 meV per unit cell energy separation between them, is found. In addition, the FM O-11-AGNR is a metal whereas the AFM O-11-AGNR is a semiconductor with a band gap of 0.17 eV. From the electronic structure trends in Fig. 2, it is also known that the NM O-11-AGNR is expected to be a metal. Furthermore, the non-planar O-11-AGNR systems have band gaps of 0.75 and 0.71 eV for $O_{SAME}$-11-AGNR and $O_{OPP}$-11-AGNR respectively. Therefore the structure found in Ref. 20 is not a ground state planar system. Rather if the band gap for an O-11-AGNR with the secondary planar geometry described in Section III.A is calculated the gap is more consistent with the result reported in Ref. 20.

As for Ref. 18, the O-AGNRs were described as planar, NM metals whose electronic trends were independent of width. Spin-polarization was not considered here. As mentioned in Section III.A, the band structure of the planar, NM O-11-AGNR is consistent with this report. As stated in the previous paragraph though, AFM and FM states for such a planar ribbon which are lower in energy than the NM one are found. Therefore it appears that magnetism in O-AGNRs remained uninvestigated in this case due to the misconception that the NM trends of the H-AGNRs would carry over in the case of a different edge passivation. It should be pointed out that bare, i.e. absence of edge passivation, AGNRs and H-passivated AGNRs do not show spin-polarization effects because of the bonding that occurs at the edges. In the bare AGNR case, the edge C atoms have a strong triple bond character with a $C_1$-$C_2$ bond length of 1.23 Å.[20] For H-passivated AGNRs, the H atom has only a single electron and therefore can only form a σ bond with the edge C atom, whereas the edge C-$p_z$ electron remains committed to the π-network throughout the AGNR. In the case of bare or H-passivated edges then there are no electrons which remain unpaired for magnetism to appear. Comparing these cases with the O-passivated systems detailed above clearly shows that there is a delicate balance between bonding at the AGNR edge and the electronic/magnetic properties which result. Such a balance is then obviously dependent upon the type of atomic species involved in passivation.

Finally there is Ref. 15 which reported the O-AGNRs as non-planar, NM semiconductors. As discussed in Section III.D, a similar result is reported where the non-planar geometry is ~2 eV per unit cell lower in energy than the planar system. This analysis also shows that such systems are indeed NM semiconductors with band gaps that decrease as a function of increasing width. Such details are not discussed in Ref. 15 though since the substitution of edge C atoms with O atoms yields a more stable structure, which is a case not considered in this paper. In addition, it should be noted that the choice of the super cell in Ref. 15 has a width that is 2× the one taken for the present calculations. The larger cell results in an O-AGNR that is considerably distorted. By doubling the size of the unit cell in Fig. 1 an energy comparison could be made between the respective non-planar geometries and the periodicity of the non-planar distortion could be investigated. After optimization it was found that the $O_{SAME}$-AGNR and $O_{OPP}$-AGNR structures are not only the lowest energy periodicity for the non-planar distortion but also 0.45 eV per super cell lower in energy than the geometry studied in Ref. 15.

## V. SUMMARY

In summary, using a first-principles DFT method the electronic and magnetic structure of the most stable planar and non-planar AGNRs whose edges have been passivated with O was investigated. In general a rich geometrical environment with this type of edge passivation, which directly correlates with the electronic and magnetic properties, is found. In the case of the planar ribbons energetically similar FM and AFM ground states that are ~0.22 eV lower in energy than the NM state are reported. DFT predicts the FM O-AGNRs to be metallic, whereas the AFM O-AGNRs fall into three families, all of which are semiconductors, except for the *3p* family which DFT predicts to be metallic. The quasiparticle corrections to the DFT band gaps using MBPT in the *GW* approximation increase the band gaps for the two semiconducting, AFM O-AGNR families. All semiconducting AFM O-AGNR families show band gaps that decrease as a function of increasing ribbon width. The magnetization in the planar systems results from the emptying of O lone-pair electrons to fill O-$p_z$ states. The magnitude of the magnetic moment per O atom as a function of nanoribbon width converges to 0.55 $\mu_B$ for both the FM and AFM states. The analysis of non-planar O-AGNRs reveals that breaking the planar symmetry results in structures whose geometry is more energetically favorable than the planar one by ~2 eV per unit cell. The distortion from a planar to a non-planar geometry is ascribed to a steric interaction between O atoms and a more efficient hybridization of atomic orbitals. In exchange, the magnetism seen in the planar systems is lost since all electrons become paired. The electronic structure of the non-planar systems also shows three families for the band gap trends, which are generally semiconductors whose energy gaps decrease as 1/width. A summary of these findings is given in Table I.

TABLE I. Summary of trends in O-AGNRs found using DFT and *GW*. For the planar magnetic systems the magnetic state is listed followed by the total magnetization and magnetization per O atom, both in $\mu_B$. Note that the following labeling conventions have been employed: P for planar, NP for non-planar, NM for nonmagnetic, FM for ferromagnetic, AFM for antiferromagnetic, M for metal, and finally S for semiconductor.

| Geometry | Magnetic ($\mu_B$) | ΔE (eV) | Family | DFT | GW |
|---|---|---|---|---|---|
| P | NM | ~2.22 | None | M | -- |
| P | FM 3.70  0.55 | ~2.00 | None | M | -- |
| P | AFM 0.00 ±0.55 | ~2.00 | 3p | M | -- |
|   |   |   | 3p+1 | S | S |
|   |   |   | 3p+2 | S | S |
| NP | NM | 0.00 | 3p[a] | S | S[b] |
|   |   |   | 3p+1[a] | S | S |
|   |   |   | 3p+2 | S | S[b] |

[a]Exceptions to the general trends: $N = 3$ and $O_{OPP}$-4-AGNR, which DFT predicts to be metallic.

In conclusion, it is reiterated that a fundamental, quantum mechanical understanding of the complicated relationship between edge passivation and the resulting electronic and magnetic properties of GNRs is vital in order to guide and interpret experimental efforts towards technologically viable applications of GNRs. This study provides a comprehensive analysis of the effects of one type of O-passivation and shows that such properties are extremely sensitive to the geometry of the nanoribbon. Furthermore, these results are used to bring consistency to the understanding of O-passivated AGNRs. For future work it is planned to explore other edge passivations to investigate the lone-pair emptying stability mechanism of double-bonded moieties in GNRs, as well as incorporate defects into the analysis.

**Acknowledgements**

This work is supported by the Interconnect Focus Center (MARCO program), State of New York, the National Science Foundation (NSF) Integrative Graduate Education and Research Traineeship (IGERT) program, Grant No. 0333314, the NSF Petascale Simulations and Analysis (PetaApps) program, Grant No. 0749140, and an anonymous gift from Rensselaer. Computing resources of the Computational Center for Nanotechnology Innovations (CCNI) at Rensselaer, partly funded by the State of New York, have been used for this work. Finally, the authors would like to thank Claus due Sinding for efforts related to this work, and SKN would like to thank Professor Mike Payne and Cavendish Laboratory for their hospitality.